# Huddling behavior in emperor penguins: dynamics of huddling

Caroline Gilbert[1*], Graham Robertson[2], Yvon Le Maho[1], Yasuhiko Naito[3], André Ancel[1]

[1] Institut Pluridisciplinaire Hubert Curien, Département Ecologie, Physiologie et Ethologie,

UMR 7178, Centre National de la Recherche Scientifique and Université Louis Pasteur.

23, rue Becquerel, 67087 Strasbourg Cedex 02, France.

Tel: +33.(0)3.88.10.69.00 / Fax: +33.(0)3.88.10.69.06

caroline.gilbert@c-strasbourg.fr

yvon.lemaho@c-strasbourg.fr

andre.ancel@c-strasbourg.fr

[2] Australian Antarctic Division.

Channel Highway, Kingston, Tasmania 7050, Australia.

Tel: +61(0)3.6232.3209 / Fax: +61(0)3.6232.3288

graham_rob@aad.gov.au

[3] National Institute of Polar Research.

1-9-10 Kaga, Itabashi, Tokyo 173-8515, Japan.

Tel: +81(3)3962.4712 / Fax: +81(3)3962.2529

bls@saturn.dti.ne.jp

* Author for correspondence: caroline.gilbert@c-strasbourg.fr

Running head: Dynamics of huddling behavior in emperor penguins

**Abstract**

GILBERT C., G. ROBERTSON, Y. LE MAHO, Y. NAITO AND A. ANCEL. Huddling behavior in emperor penguins: dynamics of huddling. PHYSIOL BEHAV 87(0) 000-000, 2006.

Although huddling was shown to be the key by which emperor penguins *(Aptenodytes forsteri)* save energy and sustain their breeding fast during the Antarctic winter, the intricacies of this social behavior have been poorly studied. We recorded abiotic variables with data loggers glued to the feathers of eight individually marked emperor penguins to investigate their thermoregulatory behavior and to estimate their "huddling time-budget" throughout the breeding season (pairing and incubation period). Contrary to the classic view, huddling episodes were discontinuous and of short and variable duration, lasting $1.6 \pm 1.7$ (SD) hours on average. Despite heterogeneous huddling groups, birds had equal access to the warmth of the huddles. Throughout the breeding season, males huddled for $38 \pm 18\%$ (SD) of their time, which raised the ambient temperature that birds were exposed to above 0°C (at average external temperatures of -17°C). As a consequence of tight huddles, ambient temperatures were above 20°C during $13 \pm 12\%$ (SD) of their huddling time. Ambient temperatures increased up to 37.5°C, close to birds' body temperature. This complex social behavior therefore enables all breeders to get a regular and equal access to an environment which allows them to save energy and successfully incubate their eggs during the Antarctic winter.

**Introduction**

The emperor penguin (*Aptenodytes forsteri*) is the only bird to breed during the severe Antarctic winter, far from the open sea or polynias where it feeds [1]. Breeding birds therefore undergo long periods of fasting. Both mates starve for about 45 days during the pairing period, while males alone take on the task of incubation, which adds another 65 days to their fast [2,3]. As a consequence, the reproductive success relies critically on the males' ability to make economic use of their body fuels. Emperor penguins are adapted to minimize heat loss [4], while maintaining their body temperature at a constant and high level [5-7]. The latter is especially important during incubation, because full embryonic development requires a temperature of about 35°C [8]. Pioneering studies have suggested that the key for the breeding success of emperor penguins is huddling [2]. Ancel et al. [9] found that field metabolic rate of huddling birds was reduced by 16% when compared with penguins that were kept in small flocks and prevented from effective huddling. The classic view is that huddles are dense formations which last for several hours [10] or even days [2]. These groups are viewed to move slowly, with the birds most exposed to the wind moving along the opposite flank of the group for protection. These huddles, formed during courtship and incubation in the colony, can be made up of more than hundreds of individuals, reaching densities of up to 10 birds per $m^2$ [2]. Kirkwood and Robertson [10] recorded ambient temperatures inside several huddles of at least 23°C, while a measurement made by Jarman [11] into a huddle suggested that the ambient temperature may reach up to 30°C.

Besides these anecdotic reports and non systematic measurements [10,11], the dynamics of huddling behavior of breeding emperor penguins in their colony during their winter fast had never been studied. Many questions remained to be answered, such as the occurrence of huddling along the nycthemeron, the duration of huddling bouts or the maximum ambient temperatures reached inside huddles. Similarly, the total time birds spend huddling during their breeding cycle (i.e. during the pairing and incubation period) had never been investigated. Huddling energetic benefits [9] also raise another question: do some birds preferentially gain from the group behavior? Emperor penguins are a social species, with no dominance hierarchy [12]: they do not defend any territory and their aggressiveness is minimal as very few struggles occur. We could therefore hypothesize that all individuals get the same benefits from huddling in order to succeed in breeding.

Consequently, the objectives of this study were (1) to characterize the occurrence and duration of huddling bouts and the microclimate created within huddles, (2) to compare individual behaviors in order to (3) estimate a "huddling time budget" for a standard breeding bird.

## 1. Materials and methods

### 1.1. Study location

The study took place at the emperor penguin colony of Pointe Géologie, Dumont d'Urville, in Adélie Land, Antarctica (66°40'S, 140°01'E). About 3000 pairs of emperor penguins make up this colony, with about 2500 incubating males during winter. The size of this colony has remained constant since the population halved in the late 1970's [13]. A meteorological station (Météo France), situated 500 meters away from the colony, provided data for wind, temperature and solar radiation, averaged every three hours.

### 1.2. Instruments and deployment protocol

In the middle of the pairing period, at the beginning of May 1998, three pairs were captured, of which males were equipped with an external time depth recorder (TDR, Mk5, Wildlife Computers, Redmond, Washington, USA, 50 g, 8 x 3.1 x 1 cm). In 2001, five pairs were captured of which the five females were equipped with an external TDR (Mk7, Wildlife Computers, 36 g, 9 x 2.4 x 1.2 cm) and an Argos-VHF transmitter (Sirtrack, Havelock North,

New-Zealand, 242 g, 13 x 5 x 3 cm). Their mates were equipped with a VHF transmitter (Sirtrack, 66 g, 10 x 1.5 x 1.5 cm).

Both mates of each pair were captured at the same time. To minimize stress, they were carefully restrained, with eyes covered. They were marked with colored strips of tape and devices were glued at the lower part of their back. To this end, a grid was worked into the feathers and covered first with Araldite (Vantico AG, Basel, Switzerland) and then a coat of mastic (resin). Loctite 401 (Henkel KGaA Technologies, Düsseldorf, Germany) was applied to the mastic and the back of the instrument, which was then glued to the mastic. All devices had been previously coated with black Tesa tape (Tesa Tape inc., Charlotte, NC, USA) to match the color of the bird feathers. Two to three Colring ties (Legrand, Limoges, France), inserted under the grid, were used to secure the instrument onto the mastic. This attachment method allowed easy removal of the instruments in the field. All devices were still securely attached to birds after the 2.5 months of this study.

Mk7 recorded external temperature (range +17°C to +42°C; resolution 0.05°C, accuracy 0.1°C) and light intensity (range 0 to 252; arbitrary unit) every 10 seconds, while Mk5 recorded temperature (range -2.5°C to +22.7°C; resolution 0.05°C, accuracy 0.1°C) and light intensity (arbitrary unit) every minute. These TDRs were calibrated in a thermostatic bath before and after deployment against a reference thermometer. The time response of this temperature sensor is estimated to be of about 5 minutes. Argos-VHF and VHF transmitters were used to locate the birds. All internal clocks were synchronized using GMT.

After the females came back from foraging at sea, on average 72 days after their departure, instruments of both males and females were removed in less than 1 minute, by cutting through the mastic, and the pairs continued with their breeding cycle. All experiments were approved by the ethics committee of the French Polar Institute.

*1.3. Data analysis*

We used the term "breeding cycle" to describe the following 2 periods: the pairing period (when both mates are in the colony) and the incubation period (when only males stay in the colony).

Light intensity was used to calculate the time a bird spent inside a huddle. A light record of zero indicated its beginning, when the bird's back was entirely covered by another bird situated behind it. A light value >0 indicated the end of the huddle. Night-time light records averaged 60 (arbitrary units), while day-time light records reached 120. Records of zero could thus be used safely to identify periods when birds were in huddles. Additional information about the density of huddles was provided by temperature sensors, as surface temperature increased when birds moved closer to each other.

Huddling patterns were classified into two categories: "tight huddles" within which surface temperature rose exponentially to above 20°C, and "huddles" within which ambient temperature never rose to 20°C. A threshold of 20°C was chosen to discriminate these 2 categories because it is the upper critical temperature of emperor penguins [6,7].

In 2001, data loggers were attached to females only. However, since males typically initiate all movements within a pair [12] and both mates huddle strictly side by side, we could also deduce information about the behavior of males from these recordings.

In order to determine which huddling strategy is chosen by breeding birds, we investigated two variables that directly determine their huddling time budget: the number of huddling bouts made per day and the durations of these episodes. Data from 1998 were used to study huddling behavior throughout the breeding cycle whereas 2001 provided additional information during the pairing period only. We counted all huddling bouts of each bird as an independent huddling event. The probability that instrumented birds were huddling within the same group is very low. Firstly, field observations showed that instrumented birds were spread out within the colony, and were therefore unlikely to be simultaneously engaged in the same huddle. Secondly, data analysis showed that huddling episodes recorded from individuals never coincided with the episodes of others.

To calculate mean durations for huddling and tight huddling bouts, we took into account only the huddling bouts lasting more than ten minutes for the winter in 1998, and more than three minutes for the winter in 2001. This was done to accommodate differences in the sampling rates of data acquisition.

Light and twilight durations at the colony site were downloaded from http://aa.usno.navy.mil/data/docs/RS_OneYear.html#formb. Twilight was defined as the time period during which the sun was between 0° to 6° below the horizon and more than 6° below the horizon defined night time. Each huddling bout overlapping on periods of twilight and light was considered as a diurnal huddling bout.

To calculate the mean number of huddling bouts made per day for each bird, we also included days when they did not huddle. Huddling bouts rarely overlapped between two days. If so, a huddling bout was attributed to the day where most of the huddling had taken place.

Time per day spent huddling and tight huddling was calculated by adding the durations of all huddling and tight huddling bouts and dividing them by 24-hours.

*1.4. Statistical analysis*

All values reported are means ± SD. All statistical tests were performed on mean values per day and showed that data were not auto-correlated. Parametric t-test (t) or non-parametric Mann-Whitney test (U), if data distribution was not normal, were used to compare periods (pairing *vs*. incubation) and years (1998 *vs*. 2001) for the various parameters (meteorological data, number of huddling and tight huddling bouts made per day, huddling and tight huddling bout duration, time spent huddling and tight huddling per day). In order to compare the percentages of time spent huddling between years, periods, and birds, data were arcsin transformed beforehand. Inter-individual comparisons were made using parametric one-way ANOVA (F) or non-parametric one-way Kruskal-Wallis ANOVA (H), if the data distribution was not normal. Post-hoc tests used in case of a significant ANOVA were either Tukey's test or Dunn's test in case of differences in sample size. The threshold of 5% was taken to determine significance. All statistical tests were performed using SigmaStat, version 2.03 (SYSTAT Software Inc., Point Richmond, California, USA).

## 2. Results

### 2.1. General results

All eight pairs studied completed their breeding cycle. All females laid one egg between the $16^{th}$ and $22^{nd}$ of May 2001 but the egg-laying dates for 1998 are not known. Out of 8 males, 7 were successful in their incubating task. One male in 2001 lost its egg around the $10^{th}$ of June, after a harsh blizzard but stayed in the colony until the $10^{th}$ of July. Males in 2001 weighed $37.6 \pm 2.4$ kg before their equipment, and $27.1 \pm 2.4$ kg after their $89.6 \pm 13.5$ days of fast, i.e. a daily weight loss of $119 \pm 13$ g per day. Males in 1998 were only weighed before their equipment in early May ($33.0 \pm 2.6$ kg). Meteorological conditions were similar between the 1998 and 2001 pairing periods (**Table 1**).

### 2.2. Huddling behavior

Between the $5^{th}$ of May and the $21^{st}$ of July 1998, 1200 huddling bouts were recorded in the 3 males, including 443 tight bouts (37%). In 2001, 305 huddling bouts were recorded in the 5 females between the $27^{th}$ of April and the $22^{nd}$ of May, including 176 tight bouts (58%). **Fig. 1** represents raw data of the huddling behavior of a pair for year 2001, over 5 five days.

#### 2.2.1. Nycthemeral distribution

Even in the middle of the winter, night is not permanent at the latitude of Pointe Géologie colony.

During the pairing periods in 1998 and 2001, 96% and 89% of the birds' huddling time occurred at night, respectively, while night accounted for 68% of the daily cycle. Only 2% (1998) and 4% (2001) of the birds' huddling time occurred during the strict light period, while 22% of the day was considered as light period.

During the 1998 incubation period, again 96% of the birds' huddling time occurred at night, while night made up 73% of the daily period. Only 2% of the birds' huddling time occurred during the light period, which made up 14% of the daily period. Thus, all birds showed a clear preference of huddling at night during both periods.

*2.2.2. Number of huddling bouts per day*

During the 1998 pairing period, the 3 males engaged in significantly more huddling bouts per day than the 5 pairs in 2001 ($5 \pm 3$ *vs*. $3 \pm 2$, respectively, U = 5514, p = 0.002). Conversely, males during the 1998 pairing period engaged in significantly less tight huddling bouts per day than the 5 pairs in 2001 ($1 \pm 1$ *vs*. $2 \pm 2$, respectively, U = 3866, p = 0.001).

During the 1998 incubation period, males engaged in more huddling bouts per day than during the pairing period ($6 \pm 2$ *vs*. $5 \pm 3$, respectively, U = 5023, p<0.001). Similarly, more tight huddling bouts per day were observed during the incubation period than during the pairing period ($3 \pm 2$ *vs*. $1 \pm 1$, respectively, U = 4243, p<0.001).

Throughout their breeding cycle in 1998, the 3 males engaged on average in $6 \pm 3$ huddling bouts per day, and $2 \pm 2$ tight huddling bouts per day. Overall, the average number of huddling and tight huddling bouts that birds engaged in per day during their breeding cycle in 1998 was similar (H = 1.888, df = 2, p = 0.383 and H = 7.492, df = 2, p = 0.024, respectively, no post-hoc differences).

During the 2001 pairing period, the 5 pairs engaged in a similar number of huddling bouts per day ($3 \pm 2$; $F_{4,88} = 0.218$, p = 0.928) and tight huddling bouts per day ($2 \pm 2$; H = 0.569, df = 4, p = 0.966). Taken together, the 3 birds in 1998 and the 5 pairs in 2001 engaged in an equivalent number of huddling and tight huddling bouts per day, though they were always observed in different huddling sub-groups within the colony.

*2.2.3. Huddling durations*

Huddling and tight huddling bouts durations ranged from a few minutes up to several hours. During the 1998 pairing period, maximum huddling and tight huddling bout duration was 7 hr 05 min and 6 hr 45 min, respectively. The respective durations for the 2001 pairing period were 9 hr 45 min and 4 hr 40 min. During the 1998 incubation period, maximum huddling and tight huddling bout duration was 11 hr 50 min and 9 hr 50 min, respectively. Including all 1998 and 2001 data, huddling bouts lasted on average $93.0 \pm 101.2$ min, and tight huddling bouts lasted $79.1 \pm 77.8$ min. Furthermore, more than 50% of huddling and tight huddling episodes lasted less than 1 hour, and about 75% of huddling episodes lasted less than 2 hours. Huddling and tight huddling bouts lasting more than 4 hours represented only 8% and 4% of all the bouts, respectively. Thus, the duration of huddling and tight huddling bouts was short and highly variable.

The duration of huddling and tight huddling bouts between years showed little difference. Huddling bouts of the 3 males during the 1998 pairing period were significantly 12% shorter than the huddling bouts of the 5 pairs during the 2001 pairing period, while the duration of their tight huddling bouts was not significantly different (**Table 2**).

Duration of huddling and tight huddling bouts for the 3 males in 1998 were not significantly different between the pairing and the incubation period (**Table 2**).

When considering the entire 1998 breeding cycle, huddling bout duration of the 3 males showed inter-individual differences, while tight huddling duration was similar (**Table 2**). These differences in huddling bout duration can be attributed to the pairing period alone. During this period, huddling bouts of male 2 were shorter than for male 1 and male 3, and tight huddling bouts of male 2 were shorter than for male 3 only (**Table 2**). In contrast, during the incubation period, the duration of huddling and tight huddling bouts was similar for the 3 males (**Table 2**).

Huddling and tight huddling bouts of the 5 pairs equipped during the 2001 pairing period did not show any inter-individual differences (**Table 2**).

*2.2.4. Temperatures reached inside huddles*

Averaged over the 1998 breeding cycle, 37% of the huddling bouts reached 20°C: 12% during the pairing period and 41% during incubation. During the 2001 pairing period, 58% of the huddling bouts induced ambient temperatures above 20°C. Thus, a great proportion of the huddles induced ambient temperatures that exceeded the upper critical temperature of 20°C for emperor penguins [6,7].

The maximum ambient temperature inside huddles recorded in our study was 37.5°C, close to the birds' deep body temperature [5-7]. Ambient temperatures during tight huddling bouts increased rapidly, occasionally reaching a peak temperature of 37.5°C. In fact, the temperatures during 38 tight huddling bouts increased from 20°C to 37.5°C within less than 2 hours (**Fig. 2**). The proportion of huddles with temperatures that reached 35°C is not negligible: out of 201 huddling bouts in 2001, during which ambient temperature was elevated above 17°C, as much as 17% rose to an ambient temperature equal to or higher than 35°C.

*2.3. Time spent huddling and tight huddling per day*

*2.3.1. Mean results*

Considering the entire breeding cycle in 1998, the 3 males spent 38 ± 18% of their daily time huddling, which included 13 ± 12% that were spent tight huddling. During the pairing and incubation period in 1998, males spent on average 29 ± 17% and 42 ± 18% of their daily time huddling, respectively. This included 6 ± 7% and 16 ± 12% spent in tight huddles during pairing and incubation period, respectively (**Fig. 3**).

During the pairing period in 2001, pairs spent $22 \pm 15\%$ of their time huddling per day, including $12 \pm 11\%$ in tight huddles. Thus, during the latter period more than 50% of the birds' huddling time was spent at ambient temperatures above 20°C (**Fig. 3**).

During the pairing period, the 3 males in 1998 spent significantly more time huddling per day than the 5 pairs in 2001 (7%; t = -2.693, p = 0.008). However, the 3 males in 1998 spent significantly less time in tight huddles per day than the 5 pairs in 2001 (6%; U = 3627.5, p = 0.008, **Fig. 3**). Hence, these results show differences between years in the time spent huddling and tight huddling per day during these two pairing periods.

In 1998 males spent more time huddling per day during the incubation period than during the pairing period (11%; t = -4.59, p<0.001). They also spent significantly more time tight huddling per day during the incubation period than during the pairing period (10%; U = 3495.5, p<0.001, **Fig. 3**). Thus, males huddled more during the incubation period than during the pairing period.

*2.3.2. Individual comparisons*

During the pairing period in 2001, each pair spent between 20 and 24% of their daily time huddling, including between 9 and 13% that was spent tight huddling. There was no significant difference between individual in the time spent huddling or tight huddling per day ($F_{4,88} = 0.175$, p = 0.951 and H = 0.609, p = 0.962, respectively).

During the breeding cycle in 1998, each male spent between 29 and 45% of its time huddling and between 11 and 14% tight huddling (**Fig. 4**). Inter-individual differences can be noticed ($F_{2,287} = 12.874$, p<0.001), as male 2 spent less time huddling per day than the other 2 males (Tukey post hoc test, p<0.05). Comparing the inter-individual behavior during the pairing period, male 1 spent significantly more time huddling than the two other males ($F_{2,57} = 27$, p<0.001, Tukey post hoc test, p<0.05). During the incubation period, however, all 3 males spent the same proportion of time huddling per day (H = 5.218, df = 2, p = 0.074).

Furthermore, no inter-individual differences were found in their tight huddling behavior (entire breeding cycle: H = 1.792, df = 2, p = 0.408; pairing period: H = 5.218, df = 2, p = 0.074; incubation period: H = 0.334, df = 2, p = 0.846). Hence, even if there were small differences in the time spent huddling per day during the pairing period, the 3 males had similar access to the warmth of the huddling groups during their breeding cycle.

By analyzing the external temperatures recorded for the 3 males, we could also determine how much time they spent at ambient temperatures above 0°C, since the temperature range of the sensors was between -2.5°C and +22.7°C. The 3 males spent on average 28 ± 11%, 43 ± 11%, and 38 ± 13% at ambient temperatures above 0°C during the pairing period, incubation, and when considering the overall breeding cycle, respectively.

## 3. Discussion

The average daily body mass loss of our five equipped birds was of 119 g per day, which is relatively low when compared with previous studies (124 [2]; 133 [14]; and 137 g per day [9]). This suggests that the effect of capture and instrumentation on birds was minor.

### 3.1. Nycthemeral distribution of huddling

The birds showed a strong preference to huddle at nighttime, during both pairing and incubation periods, which is essentially consistent with previous studies. Prévost [2] noticed that huddles during the pairing period were mostly nocturnal. Kirkwood and Robertson [10] made the same observations for females on their way to forage at sea. During incubation, Prévost [2] suggested that huddling is continuous and would last throughout the day, independently of the nycthemeral rhythm. However, we show here that huddling during both pairing and incubation periods occurs preferentially at night: 96% of the birds' huddling time occurred at night during incubation, while night made up 73% of the daily period. This might be explained by the lowered standard operative temperatures at night, when radiation from the sun is zero (0 J/cm$^2$ from 18pm to 9am local time, May-July), which in turn might prompt the birds to huddle. Huddling at night can also be linked to sleep. Miché et al. [15] showed that a day-night rhythm seems to persist for birds even at the end of May, with individual differences. Furthermore, the time spent sleeping increases as the fasting period progresses in the emperor penguin [16], which is possibly linked to an energy saving strategy. During tight huddles, we observed birds with eyes closed. We therefore suggest that huddling, and especially tight huddling, is associated with sleep.

*3.2. Huddling behavior*

*3.2.1. General discussion*

We found that equipped birds engaged in several huddling and tight huddling bouts per day. On average over the 1998 breeding cycle, males engaged in 6 ± 3 huddling bouts, and 2 ± 2 tight huddling bouts per day. Furthermore, our results show that huddling bout duration is highly variable, ranging from a few minutes up to almost 12 hours. Surprisingly, the average duration of huddling bouts is short (1.6 ± 1.7 hours, with about 75% of the huddling episodes lasting less than 2 hours). Thus, contrary to the view established from previous studies [2,10], huddling is a discontinuous phenomenon, birds being engaged in several but short lasting huddling episodes per day.

Prévost [2] reported that during incubation huddles lasted for hours or days and consisted of the same group, which was pushed by the wind. Kirkwood and Robertson [10] were the first to study the duration of huddling episodes in females during foraging trips to sea. They used data loggers attached to individuals but their sampling interval for light and temperature measurements (15 minutes) was relatively long. Their data with respect to huddling bout duration differ from our findings. In the Australian colony, huddling bouts lasted on average 12.2 ± 7.4 and 8.8 ± 8.8 hours in May and August, respectively. The longest huddle lasted 30 hours and was recorded in August. However, the data set of that study was relatively small. The differences in huddling behavior observed between the Australian and the current study could be explained by the more southerly location of the Australian colony (67°23'S and 67°28'S) and its more severe meteorological conditions and/or by the rather long sampling interval used.

One important factor, which is shown by our data and had not yet been considered before, is the heterogeneous appearance of huddling groups. Huddles are often comprised of different sub-groups, some of which consist of penguins tightly packed, while others are made up of dissociated birds (**Fig. 5**). In fact, our data describes the huddling behavior of individuals and not the overall huddling behavior of the 2500 males in the colony. This explains the short duration of huddling bouts recorded in several birds in this study, while the overall group might appear to be huddling tightly when observed from the outside [2]. This heterogeneity, that is the distinction between "huddling individuals" that are closely packed, and the "huddling group" which is composed of several groups of huddling and non-huddling birds, should therefore now be considered as important for the understanding of emperor penguin huddling behavior.

*3.2.2. Comparison between pairing and incubation*

Huddling and tight huddling bouts during the 1998 incubation period were more frequent than during the pairing period. In contrast, huddling and tight huddling bout duration during the pairing and incubation period in 1998 was not different. As a consequence, males spent in total more time huddling and tight huddling per day during the incubation period ($42 \pm 18\%$; $16 \pm 12\%$) than during the pairing period ($29 \pm 17\%$; $6 \pm 7\%$). During incubation, penguins were also subjected to higher wind speeds and lower external temperatures (**Table 1**). The difference in their huddling behavior might therefore be explained by the more severe meteorological conditions encountered during the incubation period. It might also be explained by the fact that the males' activity is reduced to a minimum when incubating the egg. During courtship, pairs have to sing, display and copulate, and are more spread out within the colony [12]. During the incubation period, however, movements of birds are restricted to their minimum, so that birds stay close together even when not huddling.

### 3.2.3. Comparison between pairing periods in 1998 and 2001

Birds during the 1998 pairing period engaged in more huddling bouts per day that lasted for a shorter duration than birds during the 2001 pairing period. The 3 males in 1998 spent significantly more time huddling per day ($29 \pm 17\%$) than the 5 pairs ($22 \pm 15\%$) in 2001. Conversely, the tight huddling bout duration was similar during both years, although birds in 2001 engaged in more tight huddling bouts per day. Consequently, the 5 pairs in 2001 spent a significantly greater time tight huddling per day ($12 \pm 11\%$) than the 3 males in 1998 ($6 \pm 7\%$).

One might expect that for a given colony, huddling behavior will change according to meteorological conditions. Our results suggest that birds behaved differently during the pairing periods of 1998 and 2001, unless the differences in huddling bout duration were due to the different types of data loggers used, their position on the birds, or the chosen sampling interval. Meteorological conditions, however, were similar during both years. The egg-laying date of the 5 females during 2001 is known, but unknown for 1998. A shift in egg-laying date and thus in the behavior of the males in 1998 might explain the observed difference in huddling behavior. Also, the study period was different between years. In 2001, we recorded data for the pairing period between April 27 and May 22, while in 1998 data were recorded between April 5 and May 31. This larger recording period might also explain the slight differences we observed in huddling behavior. For a better comparison of huddling behavior between years, we should focus on the incubation period when all other activities besides huddling are minimal. In parallel, the influence of meteorological factors on huddling occurrence should be investigated.

*3.2.4. Inter-individual comparison*

Noske [17] showed in varied Sittellas (*Daphoenositta chrysoptera*) that more dominant and older males acted as sentinels, protecting juveniles situated in the centre of a huddling group. In contrast, Calf [18] showed that in captive bronze Mannikins (*Lonchura cucullata*) dominant individuals obtained the central position within a huddle. Considering emperor penguins, we could firstly hypothesize that there could be individual differences in access to the warmth within huddles, that might rely on the amount of body fat stores of individuals and/or their breeding experience. Emperor penguins are a long-lived species: males and females begin to breed at about 5 to 6 years of age and can breed up to an age of 30-35 years [12]. Depending on its body fuel reserves, the body mass of a male at the beginning of its fast can range from about 30 to 42 kg [2]. However, a second opposed hypothesis could rely on the fact that emperor penguins are highly social, with no territory, and that major individual differences in access to the warmth within huddles could jeopardize the advantages of social thermoregulation and therefore the breeding success and survival of the birds.

Interestingly therefore, we found no differences in the number of huddling and tight huddling bouts birds engaged in per day during the periods we studied (pairing periods in 1998 and 2001 and incubation period in 1998 only). Moreover, the 3 males equipped in 1998 engaged in huddling and tight huddling bouts of equivalent duration during the incubation period. In contrast, during the pairing period, one male (male 2) behaved differently, with huddling and tight huddling bouts of shorter duration. Unfortunately again, we do not know the egg-laying date for 1998. The observed difference between individuals could be explained by the following consideration: it is possible that male 1 and 3 were already incubating, while male 2, which engaged in less huddling behavior, had not started to incubate yet. However, during the 2001 pairing period, just before the egg-laying by the 5 females, the 5 pairs showed a huddling, and tight huddling behavior, extremely similar with respect to the durations of the huddling bouts. We can therefore conclude that although the duration of huddling and tight huddling bouts was short and of variable duration within individuals, the duration of huddling bouts was similar between all birds. The high variability in huddling bout duration we observed within individuals might have also masked differences between birds.

Although birds displayed short and variable durations of huddling bouts and made several huddling bouts per day, there were no inter-individual differences in the time they spent huddling per day, except for pairing 1998. Moreover, the time spent tight huddling per day was similar between birds, despite the fact that they were always engaged in different sub-groups within the colony that consisted of 6000 birds during the pairing period and 2500 males during incubation. Birds then exhibited the same thermoregulatory behavior, and had equal access to the thermal benefits of huddles. The body mass of our studied birds at the point of instrumentation was relatively close, ranging from 35 to 41 kg in 2001, and from 30 to 35 kg in 1998. Nevertheless, a difference in 5 kg of fat could make a major difference to endure the winter fast. At a daily body mass loss of 130 g, an extra 5 kg would indeed enable birds to prolong their fast for as long as 38 days. When selecting our study animals, we probably chose 8 experienced breeders with sufficient body fat reserves. In the future, it would be most interesting to sample males across a wider body mass range or birds of different age, in order to investigate further inter-individual differences. However, our results clearly show that the time spent huddling by the equipped birds did not differ and that huddling benefits therefore seem to be equally partitioned between birds. Heterogeneity within a group and the rapid switches of individuals between a tight and a loose huddling formation are probably key factors explaining these observations.

*3.3. Temperatures reached inside huddles*

Jarman [11] and Kirwood and Robertson [10] both emphasized the effect of heating inside huddles, with ambient temperatures rising up to 20°C and beyond (temperatures exceeded 23°C, which was the upper limit for the sensors used by Kirkwood and Robertson [10]). Jarman [11] reported one record of 30°C inside a natural huddle. One striking result of our study is then the high frequency with which huddles exceeded 20°C. During the 2001 pairing period, 58% of the huddling bouts showed ambient temperatures above 20°C. Similarly, throughout the 1998 breeding cycle, 37% of the huddling bouts showed ambient temperatures above 20°C. This appears to play an important role in energy savings, though, paradoxically, these ambient temperatures are higher than the upper critical temperatures of emperor penguins.

The maximum ambient temperature recorded during our study was 37.5°C, which is close to the birds' deep body temperature [5-7]. This suggests that tight huddles could be considered as an adiabatic enclosure, at least when considering the lateral parts of the penguins' bodies. We assume that the 1.2 cm air layer [19] trapped inside the plumage of emperor penguins progressively disappears as the huddle gets tighter. An equilibrium point is then reached, at which the surface temperature of the feathers corresponds to the body temperature of the birds. An increase in feather temperature from 20°C to 37.5°C can occur in less than 2 hours, and this temperature increase is asymptotic at about 37.5°C. No heat exchange can take place between the lateral parts of birds and the environment in such a situation, reducing heat loss to its minimum. The fact that huddling bouts with temperatures above 20°C make up a great proportion of their overall huddling bouts (almost half of all huddling bouts), implies that this might represent an energy saving strategy. We can easily think that the tighter penguins huddle, the more they (1) reduce the surface area subjected to heat loss, and (2) reduce the difference in a temperature gradient. Prévost [2] reported that density inside such tight huddling sub-groups could be up to 10 birds per m$^2$. Pressure may be so high inside such groups a bird can be pushed up until it ends lying on top of its congeners (C. Gilbert, personal observation). Hence, at these densities the only exposed surfaces of birds are their head and upper back, which would enhance energy savings. In contrast, tight huddling and any increase to such unexpected high ambient temperatures (up to 37.5°C) within tight huddling groups might be risky and disadvantageous. Firstly, it could be a problem during incubation for egg development, as eggs have to be incubated at a temperature of about 35°C [8]. However, the proportion of time spent tight huddling is higher during incubation (16 ± 12%) than during the pairing period (6 ± 7%). This would mean that tight huddling is not a limitation for the egg incubation or the breeders' reproductive success. Secondly, tight huddling could lead to hyperthermia. Such temperatures are indeed well above those that induce severe heat stress in the laboratory. Pinshow et al. [7] showed that above 25°C, isolated emperor penguins had evaporative water losses 5 times higher than penguins inside their thermoneutral zone. Such a heavy water loss would be incompatible with the four-month fast of incubating males. Deep body temperature recordings of birds during tight huddling are therefore needed to investigate if tight huddling could trigger hyperthermia.

During the breeding cycle in 1998, the 3 males spent 38 ± 18% of their time huddling per day, including 13 ± 12% spent tight huddling. Furthermore, during the pairing period, the 3 males spent on average 28 ± 11% at ambient temperatures above 0°C. The equivalent values for the incubation period and the overall breeding cycle were 43 ± 11%, and 38 ± 13.0%, respectively. These daily periods spent at temperatures above 0°C (as detected from the temperature sensors) were identical to the daily periods when birds engaged in huddling, as determined by the light sensors. Although we found that throughout their breeding cycle birds spent on average only 38% of their daily time huddling, remaining in a group during the remainder of the day (even if less densely packed), must give birds a major energetic benefit by protecting them from the wind. Classically, huddling benefits are described as (1) decreasing heat loss for individuals in a huddle because of the reduced surface area to volume ratio of the huddling group [20-22] and (2) induction of a favorable microclimate by grouping [21,23]. Our data support this view. Penguins during the pairing and incubation periods can adjust their exposed body surface areas by grouping loosely or tightly. This enables them to spend a large proportion of their time at temperatures above 0°C, while external temperatures are on average -17°C. The energetic benefits that accrue from such thermoregulatory behavior of emperor penguins needs to be further investigated.

**Conclusion**

Huddling behavior of emperor penguins is a far more complex behavior than previously described. Birds make several huddling and tight huddling bouts per day, which are typically of relatively short duration. Heterogeneity within the general huddling group ensures that each individual has equal access to the warmth of huddles. Maximum ambient temperatures recorded within tight huddles reach the birds' deep body temperature. Thus, by huddling together, emperor penguins generate a "tropical" environment in one of the coldest environment on earth, which raises questions about how they manage to cope with it.

**Acknowledgements**

Field work was financially and logistically supported by the Institut Polaire Français Paul-Emile Victor (IPEV) and the Terres Australes et Antarctiques Françaises (TAAF). We thank the 48[th] and 51[st] expedition at Dumont d'Urville station for technical assistance, Météo France for meteorological data and Drs. S. Blanc and M. Enstipp for manuscript revision.

Table 1: Meteorological conditions during 1998 and 2001 for both pairing and incubation periods.

| | Wind (m.s$^{-1}$) | Temperature (°C) |
|---|---|---|
| **Pairing period (01/05 - 31/05)** n = 248 | | |
| 1998 | 7.5 ± 6.3 | -14.8 ± 4.5 |
| 2001 | 7.3 ± 5.1 | -15.0 ± 5.1 |
| *1998 vs. 2001* | *U = 61267; p = 0.821* | *t =0.396 ; p = 0.693* |
| **Incubation period (01/06 - 31/07)** n = 488 | | |
| 1998 | 8.5 ± 5.5 | -18.8 ± 5.5 |
| 2001 | 8.1 ± 6.1 | -16.3 ± 4.1 |
| *1998 vs. 2001* | *U = 245708; p = 0.096* | *U = 203459; p < 0.001* |
| *Pairing 1998 vs. incubation 1998* | *U = 82575; p = 0.001* | *U = 116501; p < 0.001* |
| **Breeding cycle (01/05 - 31/07)** n = 736 | | |
| 1998 | 8.2 ± 5.8 | -17.4 ± 5.6 |
| 2001 | 7.8 ± 5.8 | -15.8 ± 4.6 |

Table 2: Mean duration (min) of huddling (HB) and tight huddling bouts (THB) and statistical results.

| | HB | THB | HB | THB | HB | THB |
|---|---|---|---|---|---|---|
| | Pairing period | | Incubation period | | Breeding cycle | |
| **Grand mean 1998** | $81.1 \pm 89.0$ (n=295) | $67.5 \pm 63.2$ (n=70) | $97.2 \pm 111.6$ (n=905) | $78.8 \pm 81.6$ (n=373) | $93.2 \pm 106.7$ (n=1200) | $77.0 \pm 79.0$ (n=443) |
| **Grand mean 2001** | $91.9 \pm 75.7$ (n=305) | $84.2 \pm 74.6$ (n=176) | | | | |

| | HB | THB |
|---|---|---|
| Pairing 1998 *vs.* pairing 2001 | U = 81419, p < 0.001 | U = 7887, p = 0.132 |
| Pairing 1998 *vs.* incubation 1998 | U = 168898, p = 0.111 | U = 14753, p = 0.423 |
| **Inter-individual differences** | | |
| 1998 breeding cycle | H = 9.249, df = 2, p = 0.01 Dunn's post hoc test, p<0.05; male 2 < male 1 | H = 4.71, df = 2, p = 0.095 |
| 1998 pairing period | H = 22.036, df = 2, p < 0.001 Dunn's post hoc test, p < 0.05; male 2 < male 1 and male 3 | $F_{2,69}$=3.282, p=0.044 Dunn's post hoc test, p<0.05; male 2 < male 3 |
| 1998 incubation period | H = 1.751, df = 2, p = 0.417 | H = 2.745, df = 2, p = 0.254 |
| 2001 pairing period | H = 2.589, df = 4, p = 0.629 | H = 0.171, df = 4, p = 0.997 |

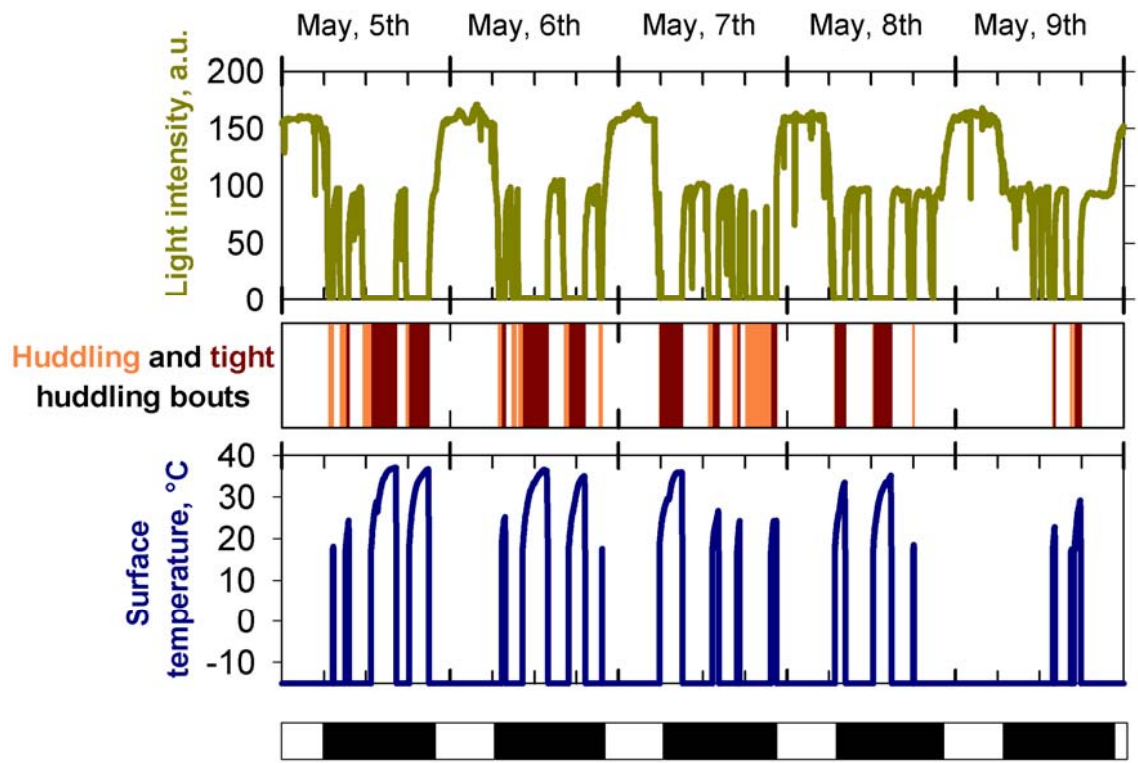

**Figure 1:** Huddling behavior of a pair on five consecutive days (May 5-9, 2001). Black boxes represent night.

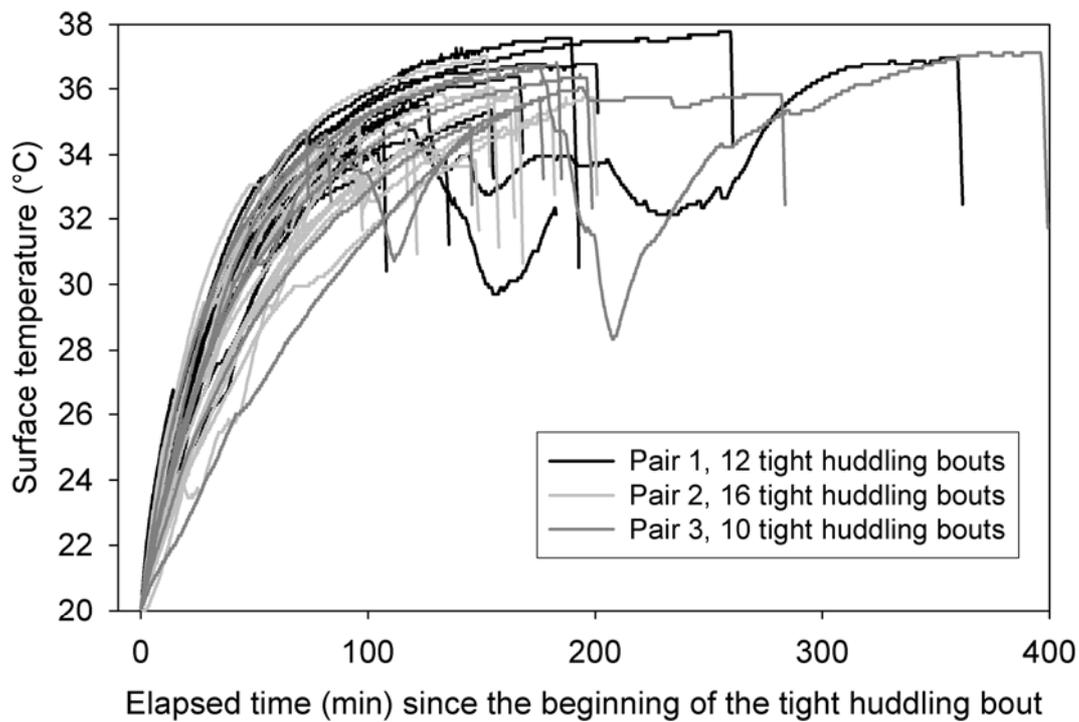

**Figure 2:** Ambient temperature increases inside tight huddling bouts for 3 pairs in 2001.

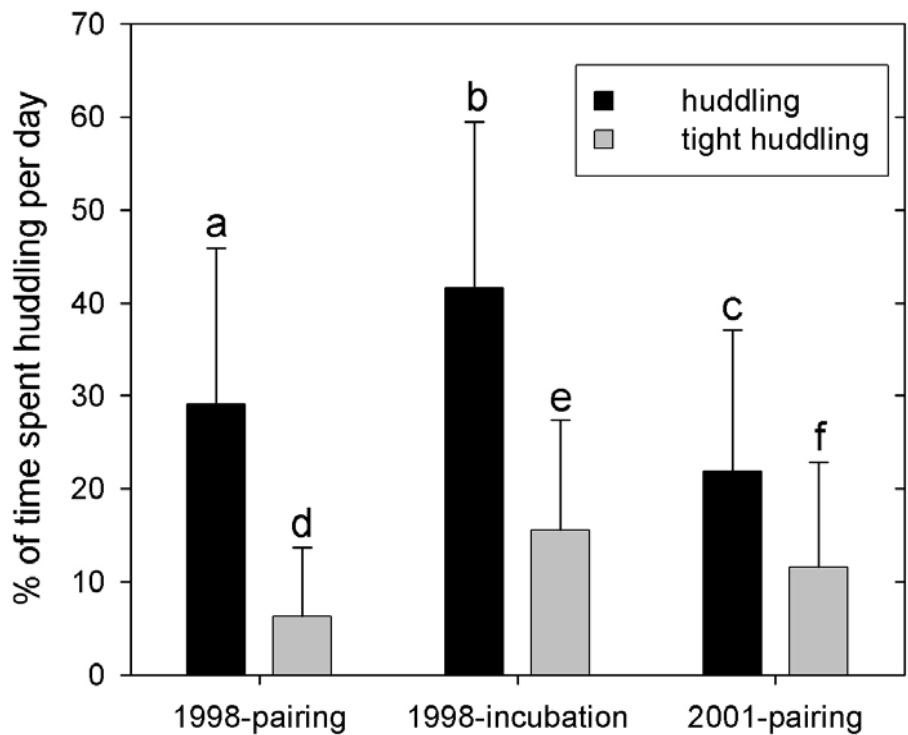

**Figure 3:** Percentage of time spent huddling and tight huddling per day during the various periods in 1998 and 2001. Values are means ± S.D. Different subscripts indicate significant differences.

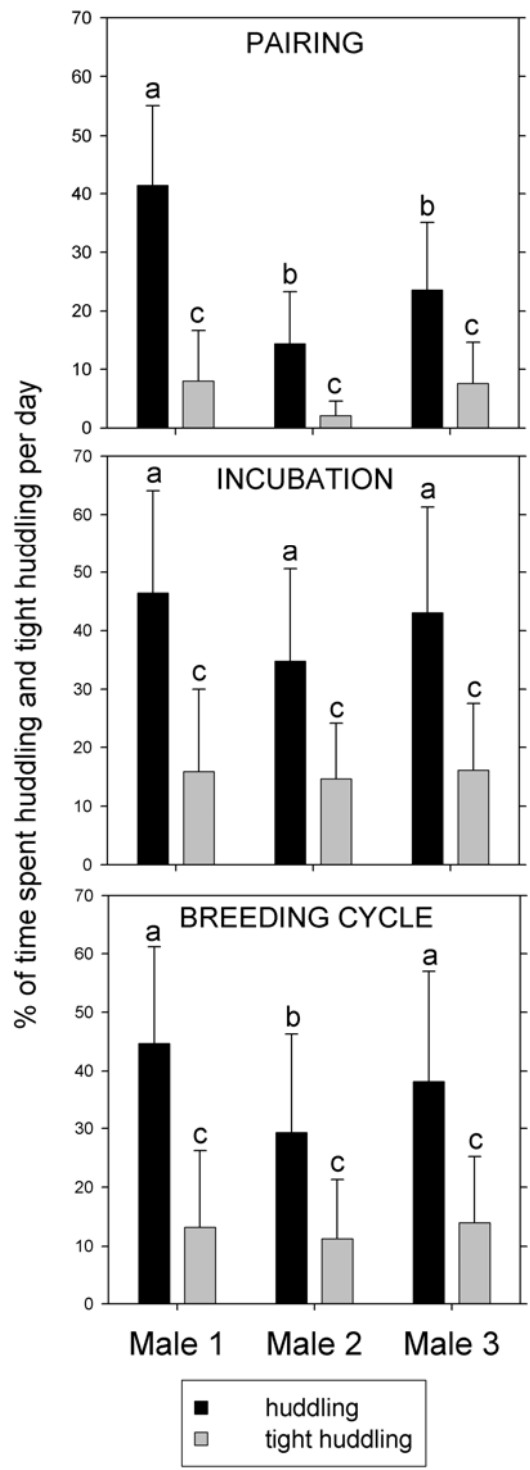

**Figure 4:** Percentage of time spent huddling and tight huddling per day by the 3 males in 1998.

Values are means ± S.D. Different subscripts indicate significant differences.

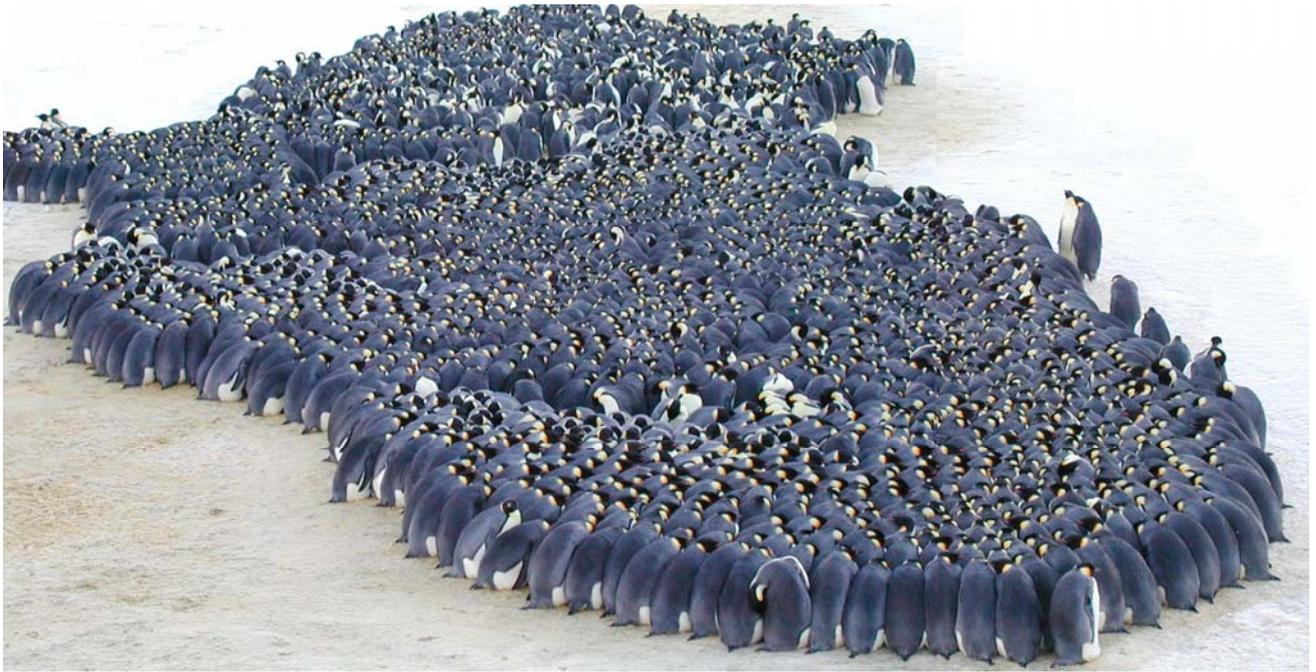

**Figure 5:** Picture of the wintering group at Pointe Géologie, which consists of about 2500 males, illustrating the heterogeneity of the huddling group.